\documentclass[twocolumn,english,aps,prd]{revtex4-1}
\usepackage{CJK}
\usepackage[T1]{fontenc}
\usepackage[left=1.5cm,top=2cm,right=1.5cm,bottom=2cm]{geometry} 
\usepackage{array}
\usepackage{textcomp}
\usepackage{multirow}
\usepackage{hhline}
\usepackage{comment}
\usepackage{adjustbox}
\usepackage[utf8]{inputenc} 
\usepackage{amsmath}
\usepackage{amsfonts}
\usepackage{amssymb}
\usepackage{cancel} 
\usepackage{youngtab}
\usepackage{xfrac}
\usepackage{graphicx}
\usepackage{subfigure} 
\usepackage{float} 

\newcommand*{\Scale}[2][4]{\scalebox{#1}{$#2$}}

\begin{document}

\title{Fermion mass hierarchy from nonuniversal abelian extensions of the Standard Model}
\author{
Carlos E. Diaz, 
S.F. Mantilla, 
R. Martinez. 
\\
{\tt\small cediazj@unal.edu.co, sfmantillas@unal.edu.co, remartinezm@unal.edu.co}
\\ \normalsize \it Departamento de F\'{i}sica, Universidad Nacional de Colombia, 
\\ \normalsize \it Ciudad Universitaria, K. 45 No. 26-85, Bogot\'a D.C., Colombia}
\date{\today}

\begin{abstract}
A nonuniversal abelian extension $\mathrm{U(1)}_{X}$  free from chiral anomalies is introduced into the Standard Model (SM), in order to evaluate its suitability in addressing the fermion mass hierarchy (FMH) by using seesaw mechanisms (SSM). In order to break the electroweak symmetry, three Higgs doublets are introduced, which give mass at tree-level to the top and bottom quarks, and the muon lepton. With an electroweak singlet scalar field, the $U(1)_{X}$ symmetry is broken and the exotic particles acquire masses. The light particles in the SM obtain their masses via SSM and Yukawa couplings differences. Active neutrino masses are generated through inverse seesaw mechanisms (ISM). Additionally, the algebraic expressions for the mixing angles for quarks and leptons are also shown in the article.

Keywords: Flavor Problem, Neutrino Physics, Extended Scalar Sectors, Beyond Standard Model, Fermion masses, Inverse seesaw Mechanism.
\end{abstract}
\maketitle 

\section{Introduction}
The current phenomenological data in high-energy physics is consistent with the existence of twelve fundamental fermions divided into quarks and leptons with their masses ranging from units of MeV to hundreds of GeV \cite{patrignani2016review}, and a large gap until thousandths of eV according to neutrino oscillation data \cite{neutrinooscilation,neutrinoexperiments}. Such particles are grouped in three chiral anomaly-free families under the gauge symmetries of the Standard Model (SM) $\mathrm{G_{SM}}=\mathrm{SU(3)}_{C}\otimes \mathrm{SU(2)}_{L}\otimes \mathrm{U(1)}_{Y}$ \cite{standardmodelfundamentals}. However, although the electroweak spontaneous symmetry breaking (SSB) accounts for the mass acquisition of fermions, it is not understood how fermions masses covers so many orders of magnitude despite there is only one vacuum expectation value (VEV) in the current SM. Moreover, the hierarchy among mixing angles of quark and lepton observed in the Cabbibo-Kobayashi-Maskawa (CKM) \cite{CKMarticles} and Pontecorvo-Maki-Nakagawa-Sakata (PMNS) matrices \cite{PMNSarticles}, respectively, has not been well understood. This issue, called \textit{fermion mass hierarchy} (FMH) \cite{fermionmasshierarchy}, is a motivation to extend the SM by adding new particles or symmetries. 

The FMH problem has been addressed from different points of view in order to achieve the simplest model beyond the SM. One of the first and most important schemes is the left-right model proposed by H. Fritzsch whose mass matrices have suited textures to understand the existence of three mass scales in the fermionic spectrum \cite{Fritzschmixings}. C. Froggat and H. Nielsen presented a model in which the heaviest fermions acquire mass through the VEV of the Higgs field, while the lightest ones get massive through radiative corrections by employing degrees of freedom heavier than the SM particles \cite{froggatt1979hierarchy}. Another remarkable way to understand the hierarchy mechanism is to assume that light neutrinos acquire their masses through radiative corrections \cite{radiativecorrections}. Similar methodologies are shown in Refs. \cite{othermethodologiesforhierarchy} using multiple scalar fields to achieve FMH. As a special case, Y. Koide addressed the model in a geometrical shape, yielding the well-known Koide formula to obtain the $\tau$ lepton mass \cite{koide1983hierarchy}, while Z. Xing analyzed the quark spectrum involving masses and mixing angles of the CKM matrix \cite{xing1996hierarchy}. Thereafter, extra dimensions were introduced in new scenarios which yielded different behaviors for each family so as their Yukawa coupling constants can account for the mass hierarchy \cite{extradimensionsforhierarchy}. Similarly, Randall-Sundrum models offer scenarios were the delocalization of fermion wavefunctions between the Planck-brane and the TeV-brane in an Anti-deSitter space could account for FMH \cite{randall1999large}.

Therefore, the confirmation of neutrino oscillations open new possibilities BSM. The smallness of neutrino masses is traditionally explained by the \textit{seesaw mechanism} (SSM) \cite{seesawmechanism}, which adds Majorana fermions $\mathcal{N}_{R}$ as right-handed neutrinos whose masses are at $\mu_{\mathcal{N}}=10^{14}$ GeV such that SM neutrinos get light masses in accordance with experimental upper limits and square mass differences, $\Delta m_{12}^{2}$ and $\Delta m_{3\ell}^{2}$, from oscillation data. However, such a shocking mass can be lower by including a second set of right-handed neutrinos $\nu_{R}$ which acquire mass at units of TeV so as the \textit{inverse SSM} (ISS) \cite{inverseseesawmechanism} can is implemented, and the Majorana mass turns out to be at units of keV $\mu_{\mathcal{N}}\sim 1 \mathrm{\,keV}$. 
Similarly, the large lepton mixings have been addressed with new methods. N. Haba and H. Murayama presented a model on neutrino masses through anarchic mass textures, i.e., without any particular structure\cite{haba2001anarchy}, and discrete symmetries such as $A_{4}$ were also employed to achieve fermion masses and mixings \cite{A4references}.

Lastly, the detection of the Higgs boson has encouraged new schemes with extended scalar sectors and gauge groups because of the existence of fundamental scalar fields in Nature. In this way, the FMH can be understood in \textit{two Higgs doublet models} (2HDM) \cite{bauer2016creating} and \textit{Next-to-Minimal 2HDMs} (N2HDM) \cite{N2HDMarticles} and N3HDM\cite{mantilla2017nonuniversal}. Thus, this article is oriented in nonuniversal abelian extensions of the SM $\mathrm{G_{SM}}\otimes \mathrm{U(1)}_{X}$ whose charges are different among families, and its symmetry breaking is ensured by a Higgs singlet $\chi$ at TeV scale. As a consequence, chiral anomalies from triangle diagrams could emerge, so it is important search for solutions of chiral anomaly equations that cancel them all, 
\begin{widetext}
\begin{eqnarray}
\label{eq:Chiral-anomalies}
\left[\mathrm{\mathrm{SU}(3)}_{C} \right]^{2} \mathrm{\mathrm{U}(1)}_{X} \rightarrow & A_{C} &= \sum_{Q}X_{Q_{L}} - \sum_{Q}X_{Q_{R}}	\nonumber	\\
\left[\mathrm{\mathrm{SU}(2)}_{L} \right]^{2} \mathrm{\mathrm{U}(1)}_{X} \rightarrow & A_{L}  &= \sum_{\ell}X_{\ell_{L}} + 3\sum_{Q}X_{Q_{L}}	\nonumber	\\
\left[\mathrm{\mathrm{U}(1)}_{Y} \right]^{2}   \mathrm{\mathrm{U}(1)}_{X} \rightarrow & A_{Y^{2}}&=
	\sum_{\ell, Q}\left[Y_{\ell_{L}}^{2}X_{\ell_{L}}+3Y_{Q_{L}}^{2}X_{Q_{L}} \right]\nonumber	
	- \sum_{\ell,Q}\left[Y_{\ell_{R}}^{2}X_{L_{R}}+3Y_{Q_{R}}^{2}X_{Q_{R}} \right]	\nonumber	\\
\mathrm{\mathrm{U}(1)}_{Y}   \left[\mathrm{\mathrm{U}(1)}_{X} \right]^{2} \rightarrow & A_{Y}&=
	\sum_{\ell, Q}\left[Y_{\ell_{L}}X_{\ell_{L}}^{2}+3Y_{Q_{L}}X_{Q_{L}}^{2} \right]\nonumber	
	- \sum_{\ell, Q}\left[Y_{\ell_{R}}X_{\ell_{R}}^{2}+3Y_{Q_{R}}X_{Q_{R}}^{2} \right]	\nonumber	\\
\left[\mathrm{\mathrm{U}(1)}_{X} \right]^{3} \rightarrow & A_{X}&=
	\sum_{\ell, Q}\left[X_{\ell_{L}}^{3}+3X_{Q_{L}}^{3} \right]\nonumber	
	- \sum_{\ell, Q}\left[X_{\ell_{R}}^{3}+3X_{Q_{R}}^{3} \right]		\nonumber	\\
\left[\mathrm{Grav} \right]^{2}   \mathrm{\mathrm{U}(1)}_{X} \rightarrow & A_{\mathrm{G}}&=
	\sum_{\ell, Q}\left[X_{\ell_{L}}+3X_{Q_{L}} \right]
	- \sum_{\ell, Q}\left[X_{\ell_{R}}+3X_{Q_{R}} \right]\nonumber		
\end{eqnarray}
\end{widetext}
Such a requirement, together with nonuniversality, is satisfied by adding new isospin singlet exotic fermions $\mathcal{T}$, $\mathcal{J}$ and $\mathcal{E}$ to the model. Furthermore, these new fermions might contribute to mass acquisition so as FMH can be obtained by avoiding unpleasant fine-tunings. Additionally, the nonuniversality in the set of $X$ charges which cancels the chiral anomalies could imply flavor violation processes such as $B^{+}\rightarrow K^{+}\ell^{+}\ell^{-}$ \cite{Bmesonanomalies} or $h\rightarrow \mu\tau$ \cite{higgsdecayanomaly}. 

The present article shows a nonuniversal abelian extension which address FMH. First of all, section \ref{sect:Suppression-squares-texture} presents the seesaw mechanism which employs the vacuum hierarchy (VH) among scalars to yield algebraic expressions of fermion masses which suggest the existence of lighter and heavier fermions than the original VEVs. After that, section \ref{sect:Models} introduces a nonuniversal model corresponding to one solution of the chiral anomaly equations \eqref{eq:Chiral-anomalies} with the corresponding analysis on the different hidden flavor symmetries and its consequences in mass acquisition mechanisms.  Then, section \ref{sect:Model-II} employs special textures to get a suited fermionic spectrum which might account for FMH. Lastly, the important features of each model are presented and compared in section \ref{sect:Conclusions} with some conclusions outlined at the end of the article. 

\section{Seesaw mechanism}
\label{sect:Suppression-squares-texture}
The majority of textures propose finite and null components of the mass matrices in order to get the suited mass eigenvalues and mixing angles. The model achieves in a natural way the fermionic mass hierarchy through  \textit{seesaw mechanisms} (SSM), with the existence of elements at two different orders of magnitude in a very special location inside the mass matrix.

The simplest example of the SSM comprises two fermions $f$ and $\mathcal{F}$ coupled by two Higgs scalars $\phi_{1,2}$. The Yukawa Lagrangian is
\begin{align}
\begin{split}
-\mathcal{L}_{Y} =\,& 
	h_{1     f     } \overline{f_{L}} {\phi}_{1} f_{R} + 
	h_{1\Scale[0.6]{\mathcal{F}}} 
	\overline{f_{L}} {\phi}_{1} \mathcal{F}_{R}+\\+\,&
	h_{2     f     } \overline{\mathcal{F}_{L}} {\phi}_{2} f_{R} + 
	h_{2\Scale[0.6]{\mathcal{F}}}
	\overline{\mathcal{F}_{L}} {\phi}_{2} \mathcal{F}_{R},
\end{split}
\end{align}
and the corresponding mass matrix after evaluating at the VEVs is
\begin{align}
M_{\mathrm{SSM}} =
\begin{pmatrix}
h_{1f} v_{1}  &	h_{1\Scale[0.6]{\mathcal{F}}} v_{1}  \\
h_{2f} v_{2}  &	h_{2\Scale[0.6]{\mathcal{F}}} v_{2}  
\end{pmatrix}.
\end{align}
The diagonalization may be done on either $MM^{\dagger}$ or $M^{\dagger}M$. Both matrices give the mass eigenvalues
\begin{align}
m_{        {f}}^{2} &\approx \frac{\left(h_{1f}h_{2\Scale[0.6]{\mathcal{F}}}-h_{2f}h_{1\Scale[0.6]{\mathcal{F}}} \right)^{2} v_{1}^{2} v_{2}^{2} }
{
\left[ \left(h_{1f}\right)^{2}+\left(h_{1\mathcal{F}}\right)^{2}\right] v_{1}^{2}+
\left[ \left(h_{2f}\right)^{2}+\left(h_{2\mathcal{F}}\right)^{2}\right] v_{2}^{2}},	\\
m_{\mathcal{F}}^{2} &\approx 
\left[ \left(h_{1f}\right)^{2}+\left(h_{1\mathcal{F}}\right)^{2}\right] v_{1}^{2}+
\left[ \left(h_{2f}\right)^{2}+\left(h_{2\mathcal{F}}\right)^{2}\right] v_{2}^{2}.
\end{align}
Regarding mixing angles, the matrix $MM^{\dagger}$ yields the left-handed mixing angle, while $M^{\dagger}M$ yields the right-handed one. They are given by
\begin{align}
\tan \theta_{L} &\approx \small{
\frac{\left[ h_{1f}h_{2f}+h_{1\Scale[0.6]{\mathcal{F}}}h_{2\Scale[0.6]{\mathcal{F}}}\right] v_{1}v_{2}}
{\left[ \left(h_{2f}\right)^{2}+\left(h_{2\Scale[0.6]{\mathcal{F}}}\right)^{2}\right] v_{2}^{2}-\left[\left(h_{1f}\right)^{2}+\left(h_{1\Scale[0.6]{\mathcal{F}}}\right)^{2}\right] v_{1}^{2}}},	\\
\tan \theta_{R} &\approx \small{
\frac{ 
h_{1f}h_{1\Scale[0.6]{\mathcal{F}}}v_{1}^{2}+
h_{2f}h_{2\Scale[0.6]{\mathcal{F}}}v_{2}^{2} }
{ \left(h_{1\Scale[0.6]{\mathcal{F}}}\right)^{2} 
v_{1}^{2}+
\left(h_{2\Scale[0.6]{\mathcal{F}}}\right)^{2}
v_{2}^{2}}}.
\end{align}

There are some remarkable features in the expressions obtained above. The first and most important is the suppression in the first eigenvalue of the matrix through the difference between Yukawa coupling constants which are all assumed at the order one, and the seesaw with the non-suppressed second eigenvalue $m_{\mathcal{F}}$ dividing such a difference. Regarding mixing angles, the left-handed angle $\theta_{L}$ is suppressed because of the VEVs, while the right-handed one does not, as it is shown below by choosing one of the two possible VHs: when $v_{1}<v_{2}$
\begin{align}
\tan \theta_{L} \approx 
\small{+
\frac{\left[ h_{1f}h_{2f}+h_{1\Scale[0.6]{\mathcal{F}}}h_{2\Scale[0.6]{\mathcal{F}}}\right]}
{\left[ \left(h_{2f}\right)^{2}+\left(h_{2\Scale[0.6]{\mathcal{F}}}\right)^{2}\right]}}
\frac{v_{1}}{v_{2}},	\quad 
\tan \theta_{R} \approx \small{
\frac{ 
  h_{2f}}
{ h_{2\Scale[0.6]{\mathcal{F}}} }}.
\end{align}
while when $v_{2}<v_{1}$ the mixing angles turn out to be
\begin{align}
\tan \theta_{L} &\approx 
\small{-
\frac{\left[ h_{1f}h_{2f}+h_{1\Scale[0.6]{\mathcal{F}}}h_{2\Scale[0.6]{\mathcal{F}}}\right]}
{\left[\left(h_{1f}\right)^{2}+\left(h_{1\Scale[0.6]{\mathcal{F}}}\right)^{2}\right]}}
\frac{v_{2}}{v_{1}},	\quad 
\tan \theta_{R} \approx \small{
\frac{ 
  h_{1f}}
{ h_{1\Scale[0.6]{\mathcal{F}}} }}.
\end{align}
In both scenarios the left-handed mixing angle gets suppressed by the specific VH between the Higgs doublets. On the contrary, the right-handed mixing angle results as the ratio among the Yukawa coupling constants of the dominant VEV. 

The following sections show the abelian extensions $\mathrm{U(1)}_{X}$ with nonuniversal sets of charges and the fermion mass acquisition in the quark and lepton sectors.

\section{Nonuniversal abelian extensions, particle content and flavor symmetries}
\label{sect:Models}

The model presented in this article is a nonuniversal abelian extension of the SM, in which a new gauge symmetry $\mathrm{U(1)}_{X}$ is added to the SM gauge group $\mathrm{G_{SM}}$. Such an extension is broken to the SM by an additional Higgs scalar singlet $\chi$ whose VEV $v_{\chi}$ lies at TeV. Then, the electroweak SSB is done by three Higgs scalar doublets $\Phi_{1,2,3}$ whose VEVs fulfill $v^{2}=v_{1}^{2}+v_{2}^{2}+v_{3}^{2}=246\mathrm{\,GeV}$ in order to get the correct masses for electroweak gauge bosons $W_{\mu}^{\pm}$ and $Z_{\mu}$. 
The complete SSB chain is
\begin{equation*}
\begin{split}
    \mathrm{SU(3)}_{C}\otimes
    \mathrm{SU(2)}_{L}\otimes 
    \mathrm{U(1)}_{Y} \otimes 
    \mathrm{U(1)}_{X} &\overset{\chi}{\longrightarrow}\\
    \mathrm{SU(3)}_{C}\otimes
    \mathrm{SU(2)}_{L}\otimes 
    \mathrm{U(1)}_{Y} &\overset{\Phi}{\longrightarrow}
    \mathrm{SU(3)}_{C}\otimes
    \mathrm{U(1)}_{Q}
\end{split}
\end{equation*}

The scalar Higgs potential of the model is given by:
\begin{equation}
\begin{split}
    V=& \mu_{1}^2\Phi_{1}^{\dagger}\Phi_{1}+\mu_{2}^2\Phi_{2}^{\dagger}\Phi_{2}+\mu_{3}^2\Phi_{3}^{\dagger}\Phi_{3}+\mu_{\chi}^2\chi^*\chi\\
    +&\mu_{12}^2(\Phi_{1}^\dagger\Phi_{2}+\Phi_{2}^\dagger\Phi_{1})+\mu_{13}^2(\Phi_{1}^\dagger\Phi_{3}+\Phi_{3}^\dagger\Phi_{1})\\
    +&\mu_{23}^2(\Phi_{2}^\dagger\Phi_{3}+\Phi_{3}^\dagger\Phi_{2})+\frac{f}{\sqrt{2}}(\Phi_{1}^\dagger\Phi_{3}\chi^*+h.c)\\
    +&\lambda_{11}(\Phi_{1}^{\dagger}\Phi_{1})^2+\lambda_{22}(\Phi_{2}^{\dagger}\Phi_{2})^2+\lambda_{33}(\Phi_{3}^{\dagger}\Phi_{3})^2+\lambda_{\chi}(\chi^{*}\chi)^2\\
    +&2\lambda_{12}(\Phi_{1}^{\dagger}\Phi_{1})(\Phi_{2}^{\dagger}\Phi_{2})+2\lambda'_{12}(\Phi_{1}^{\dagger}\Phi_{2})(\Phi_{2}^{\dagger}\Phi_{1})\\
    +&2\lambda_{13}(\Phi_{1}^{\dagger}\Phi_{1})(\Phi_{3}^{\dagger}\Phi_{3})+2\lambda'_{13}(\Phi_{1}^{\dagger}\Phi_{3})(\Phi_{3}^{\dagger}\Phi_{1})\\
    +&2\lambda_{23}(\Phi_{2}^{\dagger}\Phi_{2})(\Phi_{3}^{\dagger}\Phi_{3})+2\lambda'_{23}(\Phi_{2}^{\dagger}\Phi_{3})(\Phi_{3}^{\dagger}\Phi_{2})\\
    +&2(\lambda_{1\chi}\Phi_{1}^{\dagger}\Phi_{1}+\lambda_{2\chi}\Phi_{2}^{\dagger}\Phi_{2}+\lambda_{3\chi}\Phi_{3}^{\dagger}\Phi_{3})(\chi^{*}\chi),\\
\end{split}
\end{equation}
where the terms with $\mu_{23}^2$ break softly the discrete simmetry $\mathbb{Z}_{2}$ and the ones with $\mu_{12}^2$, $\mu_{13}^2$ the $U(1)_{X}$ gauge symmetry softly \cite{ErnestVacuum}. After symmetry breaking the minimal condition for the VEV of the scalar field $\chi$ is given by
\begin{equation}
\begin{split}
    \mu_{\chi}^2+\lambda_{\chi}v_{\chi}^2+\lambda_{1\chi}v_{1}^2+\lambda_{2\chi}v_{2}^2+\lambda_{3\chi}v_{3}^2+f\frac{v_{1}v_{3}}{v_{\chi}}=0
\end{split}    
\end{equation}
The VEV $v_{\chi}$ breaks the symmetry beyond the SM, giving masses to the exotic fermions. Therefore, since $v_{\chi}\gg v_{1}, v_{2}, v_{3}$,
\begin{equation}
    v_{\chi}^2\approx-\frac{\mu_{x}^2}{\lambda_{\chi}}
\end{equation}
The other conditions for the electroweak VEV are:
\begin{equation}
\begin{split}
    &v_{1}\left\{\mu_{1}^2+\lambda_{11}v_{1}^2+\tilde{\lambda}_{12}v_{2}^2+\tilde{\lambda}_{13}v_{3}^2+\lambda_{1\chi}v_{\chi}^2+f\frac{v_{\chi}v_{3}}{v_{1}}\right\}\\
                    &+\mu_{12}^2v_{2}+\mu_{13}^2v_{3}=0\\
    &\\
    &v_{2}\left\{\mu_{2}^2+\lambda_{22}v_{2}^2+\tilde{\lambda}_{12}v_{1}^2+\tilde{\lambda}_{23}v_{3}^2+\lambda_{2\chi}v_{\chi}^2\right\}+\mu_{12}^2v_{1}\\
                    &+\mu_{23}^2v_{3}=0\\
    &\\
    &v_{3}\left\{\mu_{3}^2+\lambda_{33}v_{3}^2+\tilde{\lambda}_{13}v_{1}^2+\tilde{\lambda}_{23}v_{2}^2+\lambda_{3\chi}v_{\chi}^2+f\frac{v_{\chi}v_{1}}{v_{3}}\right\}\\
                    &+\mu_{23}^2v_{2} +\mu_{13}^2v_{1}=0
\end{split}
\end{equation}
For $v_{1}$, $v_{2}$, $v_{3}$ the constraints are given by:
\begin{equation}\label{VEVfromSSB}
    \begin{split}
    v_{1}^2&\approx -\frac{\mu_{1}^2+\tilde{\lambda}_{12}v_{2}^2+\tilde{\lambda}_{13}v_{3}^2+\lambda_{1\chi}v_{\chi}^2+f\frac{v_{\chi}v_{3}}{v_{1}}}{\lambda_{11}}\\
    v_{2}^2&\approx \frac{\mu_{12}^4v_{1}^{2}}{\left(\mu_{2}^2+\tilde{\lambda}_{12}v_{1}^2+\tilde{\lambda}_{23}v_{3}^2+\lambda_{2\chi}v_{\chi}^2\right)^2}\\
    v_{3}^2&\approx \frac{\mu_{13}^4v_{1}^{2}}{\left(\mu_{3}^2+\tilde{\lambda}_{13}v_{1}^2+\tilde{\lambda}_{23}v_{2}^2+\lambda_{3\chi}v_{\chi}^2+f\frac{v_{\chi}v_{1}}{v_{3}}\right)^2},
    \end{split}
\end{equation}
where $\tilde{\lambda}_{ij}\equiv \lambda_{ij}+\lambda'_{ij}$. The condition for the soft symmetry breaking reads $|\mu_{1}^2|, |\mu_{2}^2|, |\mu_{3}^2|\gg |\mu_{12}^2|, |\mu_{13}^2|, |\mu_{23}^2|$, therefore $v_{1}\gg v_{2},v_{2}$. Comparing the second and third equations from (\ref{VEVfromSSB}), it can be seen that there is an additional factor $f\frac{v_{\chi}v_{1}}{v_{3}}$ in the denominator for $v_{3}^2$, which does not appear in the expression for $v_{2}^2$. Consequently, the space of parameters permits naturally the assumption of the following VH $$v_{1}^2\gg v_{2}^2\gg v_{3}^2.$$
Lastly, the scalar sector, together with the Majorana mass scale $\mu_{\mathcal{N}}$, exhibits the VH
\begin{equation}
\label{eq:Vacuum-Hierarchy}
\begin{split}
&v_{  1 } = 245.7 \mathrm{\,GeV}\sim m_{t}\sqrt{2},	\qquad	v_{\chi} = 2.5 \mathrm{\,TeV},	\\
&v_{  2 } = 12.14 \mathrm{\,GeV}\sim m_{b}\sqrt{2}, \qquad \mu_{ N} \sim 1 \mathrm{\,keV},	\\
&v_{  3 } = 250 \mathrm{\,MeV}\sim m_{\mu}\sqrt{2}. 
\end{split}
\end{equation}
This choose of VEVs of the Higgs doublets $\Phi_{1,2,3}$ plays a fundamental role on the model, because it sets the energy scale for the charm and top quarks and the muon. All other SM particles acquire masses through suppression mechanisms along with such introduced energy scales.




\begin{table}
\caption{Scalar content of the model , non-universal $X$ quantum number and $\mathbb{Z}_{2}$ parity.}
\label{tab:Bosonic-content-A-B}
\centering
\begin{tabular}{lll cll}\hline\hline 
\multirow[l]{3}{*}{
\begin{tabular}{l}
    Scalar  \\
    Doublets
\end{tabular}
}
&\multicolumn{2}{l}{}&
\multirow[l]{3}{*}{
\begin{tabular}{l}
    Scalar  \\
    Singlets
\end{tabular}
}
&\multicolumn{2}{l}{}\\ 
 &&&
 && \\ 
 &&\ $X^{\pm}$&
 &&\ $X^{\pm}$
\\ \hline\hline 
$\small{\Phi_{1}=\begin{pmatrix}\phi_{1}^{+}\\\frac{h_{1}+v_{1}+i\eta_{1}}{\sqrt{2}}\end{pmatrix}}$&&$\sfrac{+1}{3}^{+}$&
$\chi=\frac{\xi_{\chi}+v_{\chi}+i\zeta_{\chi}}{\sqrt{2}}$	&&	$\sfrac{-1}{3}^{+}$	\\
$\small{\Phi_{2}=\begin{pmatrix}\phi_{2}^{+}\\\frac{h_{2}+v_{2}+i\eta_{2}}{\sqrt{2}}\end{pmatrix}}$&&$\sfrac{+2}{3}^{-}$&
&
&		\\
$\small{\Phi_{3}=\begin{pmatrix}\phi_{3}^{+}\\\frac{h_{3}+v_{3}+i\eta_{3}}{\sqrt{2}}\end{pmatrix}}$&&$\sfrac{+2}{3}^{+}$&

	&	
&	\\  \hline\hline
\end{tabular}
\end{table}




On the other hand, the fermionic sector comprises the SM fermions (including right-handed neutrinos $\nu_{R}$) and an exotic sector composed by up-like quarks $\mathcal{T}$, down-like quarks $\mathcal{J}$, charged leptons $\mathcal{E}$ and Majorana fermions $\mathcal{N}_{R}$. The exotic sector of model has two up-like quarks $\mathcal{T}^{1,2}$, two down-like quarks $\mathcal{J}^{1,2}$, three charged leptons $\mathcal{E}^{1,2,3}$ and three Majorana masses $\mathcal{N}^{1,2,3}_{R}$. All the exotic fermions are isospin singlets, so they acquire mass through the Higgs singlet $v_{\chi}$. The addition of the exotic sector not only ensures the cancellation of chiral anomalies, which would not be canceled by only SM fermions, but also contributes to the mass acquisition mechanisms of fermions. The set of $\mathrm{U(1)}_{X}$ charges is shown in Tab. \ref{tab:Particle-content-A-B}.

\begin{table}
\caption{Particle content of the abelian extensions, non-universal $X$ quantum number and $\mathbb{Z}_{2}$ parity for the model.}
\label{tab:Particle-content-A-B}
\centering
\begin{tabular}{lll lll}\hline\hline 
\multirow[l]{3}{*}{
\begin{tabular}{l}
    Left-    \\
    Handed  
\end{tabular}
}
&\multicolumn{2}{l}{}&
\multirow[l]{3}{*}{
\begin{tabular}{l}
    Right-    \\
    Handed  
\end{tabular}
}
&\multicolumn{2}{l}{}\\ 
 &&&
 && \\ 
 &&\ $X^{\pm}$&
 &&\ $X^{\pm}$
\\ \hline\hline 
\multicolumn{6}{c}{SM Quarks}\\ \hline\hline	
\begin{tabular}{c}	
	$  q ^{1}_{L}=\begin{pmatrix}u^{1}	\\ d^{1} \end{pmatrix}_{L}$  \\
	$  q ^{2}_{L}=\begin{pmatrix}u^{2}	\\ d^{2} \end{pmatrix}_{L}$  \\
	$  q ^{3}_{L}=\begin{pmatrix}u^{3}	\\ d^{3} \end{pmatrix}_{L}$ 
\end{tabular} &
\begin{tabular}{c}
			\\
	\\		\\
	\\		\\
\end{tabular}   &
\begin{tabular}{c}
		$0^{+}$	\\
	\\	$\sfrac{+1}{3}^{-}$	\\
	\\	$\sfrac{+1}{3}^{+}$	\\
\end{tabular}   &
\begin{tabular}{c}
	$ \begin{matrix}u^{1}_{R}	\\ u^{2}_{R} \end{matrix}$  \\
	$ \begin{matrix}u^{3}_{R}	\\ d^{1}_{R} \end{matrix}$  \\
	$ \begin{matrix}d^{2}_{R}	\\ d^{3}_{R} \end{matrix}$ 
\end{tabular} &
\begin{tabular}{c}

\end{tabular} &
\begin{tabular}{c}
	$ \begin{matrix} \sfrac{+2}{3}^{+}	\\ \sfrac{+2}{3}^{-} \end{matrix}$  \\
	$ \begin{matrix} \sfrac{+2}{3}^{+}	\\ \sfrac{-2}{3}^{+} \end{matrix}$  \\
	$ \begin{matrix} \sfrac{-1}{3}^{-}	\\ \sfrac{-1}{3}^{-} \end{matrix}$ 
\end{tabular}
\\ \hline\hline 

\multicolumn{6}{c}{SM Leptons}\\ \hline\hline	
\begin{tabular}{c}	
	$\ell^{ e  }_{L}=\begin{pmatrix}\nu^{ e  }\\ e^{ e  } \end{pmatrix}_{L}$  \\
	$\ell^{\mu }_{L}=\begin{pmatrix}\nu^{\mu }\\ e^{\mu } \end{pmatrix}_{L}$  \\
	$\ell^{\tau}_{L}=\begin{pmatrix}\nu^{\tau}\\ e^{\tau} \end{pmatrix}_{L}$ 
\end{tabular} &
\begin{tabular}{c}
			\\
	\\		\\
	\\		\\
\end{tabular} &
\begin{tabular}{c}
		$\sfrac{-2}{3}^{+}$	\\
	\\	$\sfrac{-1}{3}^{-}$	\\
	\\	$-1^{+}$	\\
\end{tabular}   &
\begin{tabular}{c}
	$ \begin{matrix}\nu^{ e  }_{R}	\\ \nu^{\mu }_{R} \end{matrix}$  \\
	$ \begin{matrix}\nu^{\tau}_{R}	\\  e ^{ e  }_{R} \end{matrix}$  \\
	$ \begin{matrix} e ^{\mu }_{R}	\\  e ^{\tau}_{R} \end{matrix}$ 
\end{tabular} &
\begin{tabular}{c}
	 
\end{tabular} &
\begin{tabular}{c}
	$ \begin{matrix} \sfrac{+1}{3}^{+}	\\ 0^{-}			 \end{matrix}$  \\
	$ \begin{matrix} \sfrac{-1}{3}^{+}	\\ \sfrac{-4}{3}^{+} \end{matrix}$  \\
	$ \begin{matrix} -1^{-}				\\ \sfrac{-4}{3}^{+} \end{matrix}$ 
\end{tabular}
\\ \hline\hline 

\multicolumn{6}{c}{Non-SM Quarks}\\ \hline\hline	
\begin{tabular}{c}	
	$\mathcal{T}_{L}^{1}$	\\	$\mathcal{T}_{L}^{2}$	\\
	$\mathcal{J}_{L}^{1}$	\\	$\mathcal{J}_{L}^{2}$	
\end{tabular} &
\begin{tabular}{c}
\end{tabular}   &
\begin{tabular}{c}
	$\sfrac{+1}{3}^{-} $	\\	$+1^{-} $	\\	
	$\sfrac{-1}{3}^{+} $	\\	$ 0^{+} $	
\end{tabular}   &

\begin{tabular}{c}
	$\mathcal{T}_{R}^{1}$	\\	$\mathcal{T}_{R}^{2}$	\\
	$\mathcal{J}_{R}^{1}$	\\	$\mathcal{J}_{R}^{2}$	
\end{tabular} &
\begin{tabular}{c}
\end{tabular} &
\begin{tabular}{c}
	$\sfrac{+2}{3}^{-} $	\\	$\sfrac{+4}{3}^{-} $	\\	
	$\sfrac{-2}{3}^{+} $	\\	$\sfrac{+1}{3}^{+} $	
\end{tabular}
\\ \hline\hline

\multicolumn{6}{c}{Non-SM Leptons}\\ \hline\hline	
\begin{tabular}{c}	
    $\mathcal{E}_{L}^{1}$	\\
    $\mathcal{E}_{L}^{2}$	\\
    $\mathcal{E}_{L}^{3}$	
\end{tabular} &
\begin{tabular}{c}
\end{tabular} &
\begin{tabular}{c}
	$+1^{-}$		\\
	$-1^{+}$		\\
	$\sfrac{+5}{3}^{-}$	
\end{tabular}   &
\begin{tabular}{c}	
    $\mathcal{E}_{R}^{1}$	\\
    $\mathcal{E}_{R}^{2}$	\\
    $\mathcal{E}_{R}^{3}$	
\end{tabular} &
\begin{tabular}{c}
\end{tabular} &
\begin{tabular}{c}
	$\sfrac{+4}{3}^{-}$		\\
	$\sfrac{-4}{3}^{+}$		\\
	$\sfrac{+4}{3}^{-}$
\end{tabular}
\\	\hline\hline 

\multicolumn{3}{c}{Majorana Fermions} & 
\begin{tabular}{c}	
	$\mathcal{N}_{R}^{1,3}$	\\
	$\mathcal{N}_{R}^{2}$	
\end{tabular} &
\begin{tabular}{c}
\end{tabular} &
\begin{tabular}{c}
	$0^{+}$	\\
	$0^{-}$	
\end{tabular}
\\	\hline\hline 
\end{tabular}
\end{table}


Before acting $\mathbb{Z}_{2}$, the quark sector exhibits the global symmetry
\begin{equation}
    \mathrm{G_{Flavor}^{Q}}=
    \mathrm{SU(2)}_{q_{L}^{2,3}}\otimes
    \mathrm{SU(3)}_{u_{R}^{1,2,3}}\otimes
    \mathrm{SU(2)}_{d_{R}^{2,3}}, 
\end{equation}
while the lepton sector has the flavor symmetry
\begin{equation}
    \mathrm{G_{Flavor}^{L}}=
    \mathrm{SU(2)}_{ e _{R}^{1,3}},
\end{equation}
which does not show any universal symmetry, so it does not need breaking. 
After the action of the discrete symmetry, the quark global symmetry turns out to be
\begin{equation}
\begin{split}
    \mathrm{G_{Flavor}^{Q}}
    \overset{\mathbb{Z}_{2}}{\longrightarrow}
    \mathrm{SU(2)}_{u_{R}^{1,3}}\otimes
    \mathrm{SU(2)}_{d_{R}^{2,3}}, 
\end{split}
\end{equation}
breaking the universality in the up quark right-handed sector and ensuring the complete acquisition of masses in the fermionic sector.

The section \ref{sect:Model-II} is focused on the fermion mass acquisition, mixing angles and the different mechanisms involved to obtain the FMH.

\section{Mass hierarchy acquisition}
\label{sect:Model-II}

As it was mentioned in section III, the model lacks of flavor global symmetries. This feature implies that all fermions acquire mass at tree level. 

The hadronic sector of the model contains the SM fields with four exotic chiral quarks: two up-like quarks $\mathcal{T}^{1}$, $\mathcal{T}^{2}$ and two down-like quarks $\mathcal{J}^{1}$, $\mathcal{J}^{2}$. The leptonic sector of the model contains the SM fields with three exotic chiral charged leptons $\mathcal{E}^{1},\mathcal{E}^{2},\mathcal{E}^{3}$ and three Majorana fermions $\boldsymbol{\mathcal{N}}_{R}=(\mathcal{N}^{1},\mathcal{N}^{2},\mathcal{N}^{3}_{R})$. The non-universal quantum numbers and parities are shown in Tab.  \ref{tab:Particle-content-A-B}. 

The Yukawa Lagrangians under the symmetry $\mathrm{U(1)}_{X}\otimes \mathbb{Z}_{2}$ in the quark sector are
\begin{align}\label{yukawalagrangianup}
\begin{split}
-\mathcal{L}_{U} =\,& 
	h_{3u}^{11}\overline{q^{1}_{L}}\widetilde{\Phi}_{3}u^{1}_{R} + 
	h_{2u}^{12}\overline{q^{1}_{L}}\widetilde{\Phi}_{2}u^{2}_{R} + 
	h_{3u}^{13}\overline{q^{1}_{L}}\widetilde{\Phi}_{3}u^{3}_{R} 
	 \\+\,&
	h_{2T}^{11}\overline{q^{1}_{L}}\widetilde{\Phi}_{2}\mathcal{T}^{1}_{R}+h_{1u}^{12}\overline{q^{2}_{L}}\widetilde{\Phi}_{1}u^{2}_{R} + 
	h_{1T}^{21}\overline{q^{2}_{L}}\widetilde{\Phi}_{1}\mathcal{T}^{1}_{R}  \\+\,&
	h_{1u}^{31}\overline{q^{3}_{L}}\widetilde{\Phi}_{1}u^{1}_{R} + 
	h_{1u}^{33}\overline{q^{3}_{L}}\widetilde{\Phi}_{1}u^{3}_{R}+
	g_{\chi u}^{12}\overline{\mathcal{T}^{1}_{L}}{\chi}^{*}u^{2}_{R}  \\+\,&
	g_{\chi T}^{11}\overline{\mathcal{T}^{1}_{L}}{\chi}^{*}\mathcal{T}^{1}_{R} + 
	g_{\chi u}^{22}\overline{\mathcal{T}^{2}_{L}}{\chi}u^{2}_{R} + 
	g_{\chi T}^{21}\overline{\mathcal{T}^{2}_{L}}{\chi}\mathcal{T}^{1}_{R} \\+\,&
	g_{\chi T}^{22}\overline{\mathcal{T}^{2}_{L}}{\chi}^{*}\mathcal{T}^{2}_{R} + \mathrm{h.c.},
\end{split}
\end{align}
\begin{align}
\begin{split}
-\mathcal{L}_{D} =\,& 
	h_{3d}^{11}\overline{q^{1}_{L}}{\Phi}_{3}d^{1}_{R} + 
	h_{3J}^{13}\overline{q^{1}_{L}}{\Phi}_{3}\mathcal{J}^{1}_{R} + 
	h_{3d}^{22}\overline{q^{2}_{L}}{\Phi}_{3}d^{2}_{R} \\+\,&
	h_{3d}^{23}\overline{q^{2}_{L}}{\Phi}_{3}d^{3}_{R}+
	h_{2d}^{32}\overline{q^{3}_{L}}{\Phi}_{2}d^{2}_{R} + 
	h_{2d}^{33}\overline{q^{3}_{L}}{\Phi}_{2}d^{3}_{R} \\+\,&
	g_{\chi d}^{11}\overline{\mathcal{J}^{1}_{L}}{\chi}d^{1}_{R} + 
	g_{\chi J}^{11}\overline{\mathcal{J}^{1}_{L}}{\chi}\mathcal{J}^{1}_{R}+
	g_{\chi J}^{22}\overline{\mathcal{J}^{2}_{L}}{\chi}^{*}\mathcal{J}^{2}_{R}
	\\+\,&
	\mathrm{h.c.},
\end{split}
\end{align}
and, under the symmetry $\mathrm{U(1)}_{X}\otimes \mathbb{Z}_{2}$, the Yukawa Lagrangians in the lepton sector are
\begin{align}
\begin{split}
-\mathcal{L}_{N} =\,& 
    h_{2\nu}^{e \mu}\overline{\ell^{e}_{L}}\widetilde{\Phi}_{2}\nu^{\mu }_{R} +
    h_{1\nu}^{e\tau}\overline{\ell^{e}_{L}}\widetilde{\Phi}_{1}\nu^{\tau}_{R} + 
	h_{2\nu}^{\mu e}\overline{\ell^{\mu}_{L}}\widetilde{\Phi}_{2}\nu^{ e }_{R} \\+\,&
	h_{1\nu}^{\mu\mu}\overline{\ell^{\mu}_{L}}\widetilde{\Phi}_{1}\nu^{\mu }_{R} +
	h_{3\nu }^{\tau \tau}\overline{\ell^{\tau}_{L}}\widetilde{\Phi}_{3}\nu^{\tau}_{R} + 
	g_{\chi \mathcal{N}}^{e i} \overline{{\nu^{ e  }_{\mathcal{E}}}^{C}}{\chi}
	\mathcal{N}^{i}_{R} \\+\,&
	\overline{{\nu^{\mu }_{R}}^{C}}
	{M_{ \mathcal{N}}^{\mu i}}
	\mathcal{N}^{i}_{R}+
	g_{\chi \mathcal{N}}^{\tau i} \overline{{\nu^{\tau}_{R}}^{C}}{\chi}
	\mathcal{N}^{i}_{R} \\+\,&
	\frac{1}{2}	
	\overline{{\mathcal{N}^{i\;C}_{R}}}
	{M}_{N}^{ij}\mathcal{N}^{j}_{R} +
	\mathrm{h.c.},
\end{split}
\end{align}
\begin{align}
\begin{split}
-\mathcal{L}_{E} =\,& 
	h_{3e}^{ee}\overline{l^{e}_{L}}{\Phi}_{3}e^{e}_{R} + 
	h_{3e}^{e\tau}\overline{l^{e}_{L}}{\Phi}_{3}e^{\tau}_{R} + 
	h_{3\mathcal{E}}^{e2}\overline{l^{e}_{L}}{\Phi}_{3}\mathcal{E}^{2}_{R}  \\+\,&
	h_{3e}^{\mu \mu}\overline{l^{\mu}_{L}}{\Phi}_{3}e^{\mu}_{R}+
	h_{1e}^{\tau e}\overline{l^{\tau}_{L}}{\Phi}_{1}e^{e}_{R} + 
	h_{1e}^{\tau \tau}\overline{l^{\tau}_{L}}{\Phi}_{1}e^{\tau}_{R} \\+\,&
	h_{1\mathcal{E}}^{\tau 2}\overline{l^{\tau}_{L}}{\Phi}_{1}\mathcal{E}^{2}_{R} + 
	g_{\chi \mathcal{E}}^{11}\overline{\mathcal{E}^{1}_{L}}{\chi}^*\mathcal{E}^{1}_{R}+
	g_{\chi \mathcal{E}}^{13}\overline{\mathcal{E}^{1}_{L}}{\chi}^*\mathcal{E}^{3}_{R} \\+\,&
	g_{\chi e}^{2e}\overline{\mathcal{E}^{2}_{L}}\chi e^{e}_{R} + 
	g_{\chi e}^{2 \tau}\overline{\mathcal{E}^{2}_{L}}\chi e^{\tau}_{R}+
	g_{\chi \mathcal{E}}^{22}\overline{\mathcal{E}^{2}_{L}}\chi \mathcal{E}^{2}_{R} \\+\,&
	g_{\chi \mathcal{E}}^{31}\overline{\mathcal{E}^{3}_{L}}\chi \mathcal{E}^{1}_{R}+
	g_{\chi \mathcal{E}}^{33}\overline{\mathcal{E}^{3}_{L}}\chi \mathcal{E}^{3}_{R} + \mathrm{h.c.}
\end{split}
\end{align}
The assignation of the $U(1)_{X}$ charges for the fermions determines the X charges' configuration for the scalar fields in order to build the Yukawa Lagrangian in the fermion sector. 

For the ordinary fermionic sector, if only a Higgs doublet $\Phi_{1}$ with charge $X=1/3$ is regarded, then only two up-like quarks acquire masses, as it can be seen from the equation (\ref{yukawalagrangianup}), or in the following lines, from the mass matrix shown in equation (\ref{upmaSSMatrix}). Every down-like quark remains massless and just one of the charged leptons obtains mass, as it can be seen from the equations \eqref{eq:down-quark-masses} and \eqref{eq:Charged-Lepton-masses}, respectively. 

Additionally, when a second Higgs doublet $\Phi_{2}$ with $X=2/3$ is introduced, a down-like quark acquires mass at tree level. In the quark sector the doublets $\Phi_{1}$ and $\Phi_{2}$ give partially masses to the up-like and down-like quarks. It could be thought that there is a similarity with a Two Higgs Doublet Model type II, where the discrete symmetry $\mathbb{Z}_{2}$ is replaced by the charges X from $U(1)_{X}$. In this model a parameter $\tan(\beta)=v_{1}/v_{2}\approx m_{t}/m_{b}$ can be defined.

In order to obtain masses for the rest charged leptons, it is necessary to add a third Higgs doublet $\Phi_{3}$ with $X=2/3$, and in this situation two of this particles obtain masses. Moreover, with this addition, the rest down-like quarks and an up-like acquire masses.

The seesaw mechanisms with exotic particles also generates fine-tunning via Yukawa coupling differences from first order, which suppress some mass eigenvalues and explain how the FMH is addressed with the Higgs doublet's VEV spectrum and the singlet VEV. Furthermore, it is also possible to explain the mixing angles of quarks and leptons, as it was already shown in a similar model \cite{mantilla2017nonuniversal}.

Therefore, the X charges assignation for the scalar sector is directly derived from the structure of X charges in the fermionic sector, which gives a free anomaly $U(1)_{X}$ model. The simplest scenario to obtain mass spectrum is to assign values to $v_{1}$, $v_{2}$ and $v_{3}$ using the masses of top, bottom and muon, respectively.

Next, the Yukawa Lagrangians are evaluated at the VEVs, yielding the mass matrices whose eigenvalues, as well as their mixing angles, are shown below. The diagonalization procedure is shown in the appendix \ref{sect:Appendix-Diag}.

\subsection{Up-like quarks}
The up-like quark sector is described in the bases $\mathbf{U}$ and $\mathbf{u}$, where the former is the flavor basis while the latter is the mass basis
\begin{equation}
\begin{split}
\mathbf{U}&=(u^{1},u^{2},u^{3},\mathcal{T}^{1}),\\
\mathbf{u}&=(u, c, t, T^{1}).
\end{split}
\end{equation}
The mass term in the flavor basis turns out to be
\begin{equation}
-\mathcal{L}_{U} = \overline{\mathbf{U}_{L}} \mathbb{M}_{U} \mathbf{U}_{R}
 + {m}_{T2} \overline{\mathcal{T}_{L}^{2}} \mathcal{T}_{R}^{2} + \mathrm{h.c.},
\end{equation}
where $\mathbb{M}_{U}$ is the up-like quarks mass matrix
\begin{align}\label{upmaSSMatrix}
\mathbb{M}_{U} = \frac{1}{\sqrt{2}}
\begin{pmatrix}
h_{3u}^{11}v_{3}	&	h_{2u}^{12}v_{2}	&	h_{3u}^{13}v_{3}	&	h_{2T}^{11}v_{2}	\\	
0	&	h_{1u}^{21}v_{1}	&	0	&	h_{1T}^{21}v_{1}	\\
h_{1u}^{31}v_{1}	&	0	&	h_{1u}^{33}v_{1}	&	0	\\
0	&	g_{\chi u}^{12}v_{\chi}	&	0	&	g_{\chi T}^{11}v_{\chi}	\\
\end{pmatrix}
\end{align}

Since the determinant of $\mathbb{M}_{U}$ is non-vanishing, all up-like quarks acquire mass. Then, the mass eigenvalues corresponding with the SM quark masses are
\begin{equation}
\label{eq:Up-Quarks-masses}
\begin{split}
m_{u}^{2} &\approx \frac{(h_{3u}^{11}h_{1u}^{33}-h_{3u}^{13}h_{1u}^{31})^{2}}{(h_{1u}^{31})^{2}+(h_{1u}^{33})^{2}}\frac{v_{3}^{2}}{2},\\
m_{c}^{2} &\approx  \frac{\left(h_{1u}^{21}g_{\chi u}^{12}-h_{1T}^{21}g_{\chi T}^{11}\right)^{2}}{(g_{\chi u}^{12})^{2}+(g_{\chi T}^{11})^{2}}
\frac{v_{1}^{2}}{2},    \\
m_{t}^{2} &\approx \left[(h_{1u}^{31})^{2}+(h_{1u}^{33})^{2}\right)]
\frac{v_{1}^{2}}{2},	\\
\end{split}
\end{equation}
while the masses of the exotic up-like quarks are
\begin{equation}
\label{eq:Exotic-Up-Quarks-masses}
\begin{split}
m_{T1}^{2}&\approx \left[(g_{\chi u}^{12})^{2}+(g_{\chi T}^{11})^{2}\right]\frac{v_{\chi}^{2}}{2}\\  
m_{T2}^{2}&\approx (g_{\chi T}^{22})^{2}\frac{v_{\chi}^{2}}{2}.
\end{split}
\end{equation}
The corresponding left-handed rotation matrix can be expressed by
\begin{equation}
\mathbb{V}^{U}_{L} = \mathbb{V}^{U}_{L,\mathrm{SS}}\mathbb{V}^{U}_{L,\mathrm{B}}, 
\end{equation}
where the seesaw angle is
\begin{equation}
\Theta^{U\dagger}_{L} = 
\begin{pmatrix}
 \dfrac{(h_{2u}^{12}g_{\chi u}^{12}+h_{2T}^{12}g_{\chi T}^{11})}{(g_{\chi u}^{12})^{2}+(g_{\chi T}^{11})^{2}}\dfrac{v_{2}}{v_{\chi}}e^{i(b_{U}-e_{U})} \\
 \dfrac{(h_{1u}^{21}g_{\chi u}^{12}+h_{1T}^{21}g_{\chi T}^{11})}{(g_{\chi u}^{12})^{2}+(g_{\chi T}^{11})^{2}}\dfrac{v_{1}}{v_{\chi}}e^{i(c_{U}-e_{U})} \\
 0 \\
\end{pmatrix}
\end{equation}
while $\mathbb{V}^{U}_{L,\mathrm{B}}$ diagonalizes only the SM-up quarks. Its angles are given by
\begin{equation}
\begin{split}
\tan \theta_{13}^{U,L} &\approx 
	\frac{h_{3u}^{11}h_{1u}^{31}+h_{3u}^{13}h_{1u}^{33}}{(h_{1u}^{31})^{2}+(h_{1u}^{33})^{2}}
	\frac{v_{3}}{v_{1}}
	e^{i(a_{U}-d_{U})}
	\\
\tan \theta_{23}^{U,L} &\approx 0
\\
\tan \theta_{12}^{U,L} &\approx 
	\frac{(h_{2T}^{11}g_{\chi u}^{12}-h_{2u}^{12}g_{\chi T}^{11})}{(h_{2T}^{21}g_{\chi u}^{12}-h_{2u}^{22}g_{\chi T}^{11})}\frac{v_{2}}{v_{1}}e^{i(b_{U}-c_{U})}
\end{split}
\end{equation}

The exotic species $T^{1}$ and $T^{2}$ got masses through $v_{\chi}$ at units of TeV. The SM $t$ quark has acquired mass with $v_{1}$ without any suppression, so its mass remains at the scale of $v_{1}$, hundreds of GeV. On the contrary, the $c$ quark have acquired mass with $v_{1}$ and $v_{2}$ but 
yielding the suppressed mass of the $c$ quark because of the existance of a SSM together with the $T^{1}$ exotic quark.
Finally, the $u$ quark has acquired mass through $v_{3}$ with a similar SSM as for $c$ quark, but with the $t$ quark instead of $T^{1}$.

Consequently, the mass of the $u$ quark gets suppressed by the top quark $t$.

\subsection{Down-like quarks}
The down-like quarks are described in the bases $\mathbf{D}$ and $\mathbf{d}$, where the former is the flavor basis while the latter is the mass basis
\begin{equation}
\begin{split}
\mathbf{D}&=(d^{1},d^{2},d^{3},\mathcal{J}^{1}),\\
\mathbf{d}&=(d, s, b, J^{1}).
\end{split}
\end{equation}
The mass term in the flavor basis is
\begin{equation}
-\mathcal{L}_{D} = \overline{\mathbf{D}_{L}} \mathbb{M}_{D} \mathbf{D}_{R}
 + {m}_{J2} \overline{\mathcal{J}_{L}^{2}} \mathcal{J}_{R}^{2} + \mathrm{h.c.},
\end{equation}
where $\mathbb{M}_{D}$ turns out to be 
\begin{align}
\mathbb{M}_{D} = \frac{1}{\sqrt{2}}
\begin{pmatrix}
h_{3d}^{11}v_{3}	&	0	&	0	&	h_{3\mathcal{J}}^{11}v_{3}	\\
0	&	h_{3d}^{22}v_{3} 	&	h_{3d}^{23}v_{3}	&	0	\\
0	&	h_{2d}^{32}v_{2}	&	h_{2d}^{33}v_{2}	&	0	\\
g_{\chi d}^{11}v_{\chi}	& 0	
&	0   
&
g_{\chi \mathcal{J}}^{11}v_{\chi}	\\
\end{pmatrix}
\end{align}

Thus, the mass eigenvalues of the SM quarks are
\begin{equation}\label{eq:down-quark-masses}
\begin{split}
m_{d}^{2} &\approx \frac{(h_{3\mathcal{J}}^{11}g_{\chi d}^{11}-h_{3d}^{11}g_{\chi \mathcal{J}}^{11})^2}{(g_{\chi d}^{11})^2+(g_{\chi \mathcal{J}}^{11})^2}\frac{v_{3}^2}{2},\\
m_{s}^{2} &\approx \frac{(h_{3d}^{23}h_{2 d}^{32}-h_{3d}^{22}h_{2 d}^{33})^2}{(h_{2 d}^{32})^2+(h_{2 d}^{33})^2}\frac{v_{3}^2}{2},\\
m_{b}^{2} &\approx \left[(h_{2 d}^{32})^2+(h_{2 d}^{33})^2\right]\frac{v_{2}^{2}}{2}   
\end{split}
\end{equation}
and the masses of the exotic species are given by
\begin{equation}
\begin{split}
m_{J1}^{2}&\approx \left[(g_{\chi d}^{11})^2+(g_{\chi \mathcal{J}}^{11})^2\right]\frac{v_{\chi}^2}{2}\\
m_{J2}^{2}&\approx (g_{\chi \mathcal{J}}^{22})^{2}\frac{v_{\chi}^{2}}{2}.
\end{split}
\end{equation}

The corresponding left-handed rotation matrix is
\begin{equation}
\mathbb{V}^{D}_{L} = \mathbb{V}^{D}_{L,\mathrm{SS}}\mathbb{V}^{D}_{L,\mathrm{B}}, 
\end{equation}
where the seesaw angle which rotates out species $J^{1,2}$ is
\begin{equation}
\Theta^{D\dagger}_{L} = 
\begin{pmatrix}
\dfrac{h_{3d}^{11}g_{\chi d}^{11}+h_{3\mathcal{J}}^{11}g_{\chi \mathcal{J}}^{11}}{(g_{\chi d}^{11})^2+(g_{\chi \mathcal{J}}^{11})^2}\dfrac{ v_{3} }{ v_{\chi}}
 \\0
\\0
\end{pmatrix},
\end{equation}
and the SM angles of $\mathbb{V}^{D}_{L,\mathrm{B}}$ are given by
\begin{equation}
\begin{split}
\tan \theta_{13}^{D,L} &\approx 
	0,	\\
\tan \theta_{23}^{D,L} &\approx 
	\frac{(h_{3d}^{22}h_{2d}^{32}+h_{3d}^{23}h_{2d}^{33})}{(h_{2d}^{23})^2+(h_{2d}^{33})^2}\frac{v_{3}}{v_{2}},
		\\
\tan \theta_{12}^{D,L} &\approx 0 
\end{split}
\end{equation}
The heaviest quarks $J^{1}$ and $J^{2}$ acquired mass at TeV scale due to $v_{\chi}$, while the $b$ quark obtained its mass through $v_{2}$ at units of GeV. The $s$ quark has acquired its mass through $v_{3}$ at hundreds of MeV with the corresponding SSM with the bottom quark $b$.  

Similarly, the quark $d$ got its mass through the SSM with the exotic species $J^{1}$.

\subsection{Neutral leptons}
Neutrinos involve Dirac and Majorana masses in their Yukawa Lagrangian. Since $\mathcal{N}_{R}^{i}$ are Majorana fermions, the bases are chiral and the mass basis describes Majorana neutrinos. The flavor and mass bases are, respectively,
\begin{equation}
\begin{split}
\mathbf{N}_{L}&=(\nu^{e,\mu,\tau}_{L},(\nu^{e,\mu,\tau}_{R})^{C},(\mathcal{N}^{e,\mu,\tau}_{R})^{C}),\\
\mathbf{n}_{L}&=(\nu^{1,2,3}_{L},(N^{1,2,3}_{R})^{\mathrm{C}},(\tilde{N}^{1,2,3}_{R})^{\mathrm{C}}).
\end{split}
\end{equation}
The mass term expressed in the flavor basis is
\begin{equation}
-\mathcal{L}_{N} = \frac{1}{2} \overline{\mathbf{N}_{L}^{C}} \mathbb{M}_{N} \mathbf{N}_{L},
\end{equation}
where the mass matrix has the following block structure
\begin{equation}
\mathbb{M}_{N} = 
\left(\begin{array}{c c c}
0	&	\mathcal{M}_{\nu}^{\mathrm{T}}	&	0	\\
\mathcal{M}_{\nu}	&	0	&	\mathcal{M}_{\mathcal{N}}^{\mathrm{T}}	\\
0	&	\mathcal{M}_{\mathcal{N}}	&	M_{\mathcal{N}}
\end{array}\right),
\end{equation}
with $\mathcal{M}_{\nu}$ as the Dirac mass matrix between $\nu_{L}$ and $\nu_{R}$
\begin{equation}
\label{eq:m_nu_original_parameters}
\mathcal{M}_{\nu} = 
\frac{1}{\sqrt{2}}
\left(\begin{matrix}
0	&	h_{2\nu}^{e\mu} v_{2}	&	h_{1\nu}^{e\tau} v_{1}	\\
h_{2\nu}^{\mu e} v_{2}	&	h_{1\nu}^{\mu\mu} v_{1}	&	0	\\
0	&	0	&	h_{3\nu}^{\tau\tau} v_{3}	
\end{matrix}\right),
\end{equation}

$\mathcal{M}_{N}$ the Dirac mass matrix between $\nu_{R}^{C}$ and $\mathcal{N}_{R}$
\begin{equation}
\mathcal{M}_{N} = 
\frac{v_{\chi}}{\sqrt{2}}
\left(\begin{matrix}
g_{\chi \mathcal{N}}^{e 1}	&	g_{\chi \mathcal{N}}^{e 2}	&	g_{\chi \mathcal{N}}^{e 3}	\\
g_{ \mathcal{N}}^{\mu 1}	&	g_{ \mathcal{N}}^{\mu 2}	&	g_{ \mathcal{N}}^{\mu 3}	\\
g_{\chi \mathcal{N}}^{\tau 1}	&	g_{\chi \mathcal{N}}^{\tau 2}	&	g_{\chi \mathcal{N}}^{\tau 3}
\end{matrix}\right),
\end{equation}
where $g_{ \mathcal{N}}^{\mu i}=\sqrt{2} M_{ \mathcal{N}}^{\mu i}/v_{\chi}$, and $M_{\mathcal{N}}=\mathbb{G}_{N}\mu_{\mathcal{N}}$ is the Majorana mass of $\mathcal{N}_{R}$. 

By employing the inverse SSM because of the VH in eq. \eqref{eq:Vacuum-Hierarchy}, it is found that 
\begin{equation}
\left(\mathbb{V}_{L,\mathrm{SS}}^{N}\right)^{\dagger}\mathbb{M}_{N} \mathbb{V}_{L,\mathrm{SS}}^{N} = 
\begin{pmatrix}
m_{\nu}	&	0	&	0	\\
0	&	m_{N}	&	0	\\
0	&	0	&	m_{\tilde{N}}
\end{pmatrix}
\end{equation}
where the new $3\times 3$ blocks are\cite{catano2012neutrino}
\begin{equation}
\label{eq:Neutrino-block-mass-matrices}
\begin{split}
m_{\nu} &=	\mathcal{M}_{\nu}^{\mathrm{T}} \left( \mathcal{M}_{\mathcal{N}}^{\mathrm{T}} \right)^{-1} M_{\mathcal{N}} \left( \mathcal{M}_{\mathcal{N}} \right)^{-1} \mathcal{M}_{\nu},	\\
M_{       N } &\approx \mathcal{M}_{\mathcal{N}}-{M}_{\mathcal{N}},	\quad	
M_{\tilde{N}}\approx \mathcal{M}_{\mathcal{N}}+{M}_{\mathcal{N}}.
\end{split}
\end{equation}

It was assumed $\mathcal{M}_{N}$ diagonal and
\begin{equation}
\mathbb{G}_{N} = 
\begin{pmatrix}
G_{N1}	&	G_{N4}	&	0	\\
G_{N4}	&	G_{N2}	&	0	\\
	0	&		0	&	G_{N3}
\end{pmatrix}
\end{equation}
so as it can yield the adequate mixing angles to fit PMNS matrix. By rejecting terms proportional to $v_{3}$ in $m_{\nu}$, the neutrino $\nu^{1}_{L}$ turns out to be massless, the masses of the other two neutrinos are 
\begin{equation}
\begin{split}
m_{\nu2}  &\approx \frac{(h_{2\nu}^{\mu e})^2 G_{N2} }{(g_{ \mathcal{N}}^{\mu 2})^2}\frac{\mu_{N}v_{1}^{2}}{v_{\chi}^{2}}
,	\\ 
m_{\nu3}  &\approx \frac{(h_{2\nu}^{e\mu})^2 G_{N1} }{(g_{\chi\mathcal{N}}^{e 1})^2}\frac{\mu_{N}v_{1}^{2}}{v_{\chi}^{2}}
\end{split}
\end{equation}
and the masses of the exotic species are
\begin{equation}
\begin{split}
m_{N^{1}_{R}} &= \frac{g_{\chi \mathcal{N}}^{e1}v_{\chi}}{\sqrt{2}} \pm \frac{G_{N1}\mu_{N}}{2},	
	\\
m_{N^{2}_{R}} &= \frac{g_{\mathcal{N}}^{\mu 2} v_{\chi}}{\sqrt{2}} \pm \frac{G_{N2}\mu_{N}}{2},	
	\\
m_{N^{3}_{R}} &= \frac{g_{\chi\mathcal{N}}^{\tau 3}v_{\chi}}{\sqrt{2}} \pm \frac{G_{N3}\mu_{N}}{2}.	
\end{split}
\end{equation}
The left-handed rotation matrix can be expressed by
\begin{equation}
\mathbb{V}^{N}_{L} = \mathbb{V}^{N}_{L,\mathrm{SS}}\mathbb{V}^{N}_{L,\mathrm{B}},
\end{equation}
where the seesaw angle is
\begin{equation}
\Theta^{N\dagger}_{L} =
\left(
\begin{array}{ccc}
 \frac{h_{2\nu}^{\mu e} G_{\text{N4}} v_2\mu_{N}}{g_{\chi\mathcal{N}}^{e1} g_{\mathcal{N}}^{\mu 2} v_{\chi}^{2}}
  & \frac{h_{1\nu}^{\mu\mu} G_{\text{N4}} v_1\mu_{N}}{g_{\chi\mathcal{N}}^{e1} g_{\mathcal{N}}^{\mu 2}v_{\chi}^{2}}
  & \frac{h_{1\nu}^{e\tau} G_{\text{N1}} v_1\mu_{N}}{(g_{\chi\mathcal{N}}^{e1})^2 v_{\chi}^{2}} \\
 \frac{h_{2\nu}^{\mu e} G_{\text{N2}} v_2\mu_{N}}{(g_{\mathcal{N}}^{\mu 2})^2v_{\chi}^{2}}
  & \frac{h_{1\nu}^{\mu\mu} G_{\text{N2}} v_1\mu_{N}}{(g_{\mathcal{N}}^{\mu 2})^2v_{\chi}^{2}}
   & \frac{h_{1\nu}^{e\tau} G_{\text{N4}} v_1\mu_{N}}{g_{\chi\mathcal{N}}^{e1} g_{\mathcal{N}}^{\mu 2} v_{\chi}^{2}} \\
 0 & 0 & \frac{h_{3\nu}^{\tau\tau} G_{\text{N3}} v_3\mu_{N}}{(g_{\chi\mathcal{N}}^{\tau 3})^2 v_{\chi}^{2}}
\end{array}
\right)
\end{equation}
and $\mathbb{V}^{E}_{L,\mathrm{SM}}$, contained in the block-diagonal mixing matrix $\mathbb{V}^{E}_{L,\mathrm{B}}$ after rotating out the heavy species has the angles
\begin{equation}
\label{eq:Neutrino-SM-Rotation-angles}
\begin{split}
\tan \theta_{13}^{N,L} &\approx 
	\frac{h_{2\nu}^{e\mu}h_{2\nu}^{\mu e}v_{2}^{2}}{h_{1\nu}^{e\tau}h_{1\nu}^{\mu\mu}v_{1}^{2}},
	\\
\tan \theta_{23}^{N,L} &\approx 
	\frac{h_{1\nu}^{e\tau} h_{2\nu}^{\mu e} g_{\chi\mathcal{N}}^{e1} g_{\mathcal{N}}^{\mu 2} G_{N4}}
	{(h_{1\nu}^{e\tau})^2 (g_{\mathcal{N}}^{\mu 2})^2 G_{N1} - (h_{1\nu}^{\mu\mu})^2 (g_{\chi\mathcal{N}}^{e1})^2 G_{N2}}\frac{v_{2}}{v_{1}}
	\\
\tan \theta_{12}^{N,L} &\approx 
	\frac{h_{2\nu}^{\mu e}v_{2}}{h_{1\nu}^{\mu\mu}v_{1}}.
\end{split}
\end{equation}

\subsection{Charged leptons}
The charged leptons are described in the bases $\mathbf{E}$ and $\mathbf{e}$, where the former is the flavor basis while the latter is the mass basis
\begin{equation}
\begin{split}
\mathbf{E}&=(e^{e},e^{\mu} , e^{\tau}, \mathcal{E}^{1}, \mathcal{E}^{2}, \mathcal{E}^{3}),\\
\mathbf{e}&=(e, \mu, \tau, E^{1}, E^{2}, E^{3}).
\end{split}
\end{equation}
The mass term obtained from the Yukawa Lagrangian is
\begin{equation}
\label{eq:Electron-mass-matrix}
-\mathcal{L}_{E} = \overline{\mathbf{E}_{L}}\mathbb{M}_{E} \mathbf{E}_{R} + \mathrm{h.c.}
\end{equation}
where $\mathbb{M}_{E}$ turns out to be 
\begin{widetext}
\begin{align}
\mathbb{M}_{E} = \frac{1}{\sqrt{2}}
\begin{pmatrix}
h_{3e}^{ee}v_{3}	&	0	&	h_{3e}^{e\tau}v_{3}	&	0	&	h_{3\mathcal{E}}^{e2}3v_{3}	&	0	\\	
0	&	h_{3e}^{\mu\mu}v_{3}	&	0	&	0	&	0	&	0	\\
h_{1e}^{\tau e}v_{1}	&	0	&	h_{1e}^{\tau \tau}v_{1}	&	0	&	h_{1\mathcal{E}}^{\tau 2}v_{1}	&	0	\\
0	&	0	&	0	&	g_{\chi \mathcal{E}}^{11}v_{\chi}	&	0	&	g_{\chi \mathcal{E}}^{13}v_{\chi}	\\
g_{\chi e}^{2e}v_{\chi}	&	0	&	g_{\chi e}^{2\tau}v_{\chi}	&	0	&	g_{\chi \mathcal{E}}^{22}v_{\chi}	&	0	\\
0	&	0	&	0	&	g_{\chi \mathcal{E}}^{31}v_{\chi}	&	0	&	g_{\chi \mathcal{E}}^{33}v_{\chi}
\end{pmatrix}
\end{align}

The determinant of $\mathbb{M}_{E}$ is non-vanishing ensuring that all charged leptons acquire mass. Thus, the eigenvalues of the mass matrix yields the masses of the SM leptons
\begin{equation}
\label{eq:Charged-Lepton-masses}
\begin{split}
m_{ e  }^{2} &\approx \frac{\left[(h_{3\mathcal{E}}^{e2}h_{1e}^{\tau\tau}-h_{3e}^{e\tau}h_{1\mathcal{E}}^{\tau 2})g_{\chi e}^{2e}-(h_{3\mathcal{E}}^{e2}h_{1e}^{\tau e}-h_{3e}^{ee}h_{1\mathcal{E}}^{\tau 2})g_{\chi e}^{2\tau}+(h_{3e}^{e\tau}h_{1e}^{\tau e}-h_{3e}^{ee}h_{1e}^{\tau \tau})g_{\chi \mathcal{E}}^{22}\right]^2}{(h_{1e}^{\tau\tau}g_{\chi e}^{2e}-h_{1e}^{\tau e}g_{\chi e}^{2\tau})^2+(h_{1\mathcal{E}}^{\tau 2}g_{\chi e}^{2e}-h_{1e}^{\tau e}g_{\chi \mathcal{E}}^{22})^2+(h_{1\mathcal{E}}^{\tau 2}g_{\chi e}^{2\tau}-h_{1e}^{\tau \tau}g_{\chi\mathcal{E}}^{22})^2}\frac{v_{3}^2}{2},\\
m_{\mu }^{2} &\approx (h_{3e}^{\mu\mu})^{2}\frac{v_{3}^{2}}{2},\\
m_{\tau}^{2} &\approx \frac{(h_{1e}^{\tau\tau}g_{\chi e}^{2e}-h_{1e}^{\tau e}g_{\chi e}^{2\tau})^2+(h_{1\mathcal{E}}^{\tau 2}g_{\chi e}^{2e}-h_{1e}^{\tau e}g_{\chi \mathcal{E}}^{22})^2+(h_{1\mathcal{E}}^{\tau 2}g_{\chi e}^{2\tau}-h_{1e}^{\tau \tau}g_{\chi\mathcal{E}}^{22})^2}{(g_{\chi e}^{2e})^2+(g_{\chi e}^{2\tau})^2+(g_{\chi \mathcal{E}}^{22})^2}\frac{v_{1}^2}{2},
\end{split}
\end{equation}
and the masses of the new exotic charged leptons
\begin{equation}
\label{eq:Exotic-Charged-Lepton-masses}
\begin{split}
m_{E1}^{2}&\approx ,\left[(g_{\chi\mathcal{E}}^{11})^2+(g_{\chi\mathcal{E}}^{13})^2\right]\frac{v_{\chi}^{2}}{2}\\
m_{E2}^{2}&\approx  \left[(g_{\chi e}^{2e})^2+(g_{\chi e}^{2\tau})^2+(g_{\chi \mathcal{E}}^{22})^2\right]\frac{v_{\chi}^{2}}{2},\\
m_{E3}^{2}&\approx \left[(g_{\chi\mathcal{E}}^{31})^2+(g_{\chi\mathcal{E}}^{33})^2\right]\frac{v_{\chi}^{2}}{2}.  
\end{split}
\end{equation}
The left-handed rotation matrix can be expressed by
\begin{equation}
\mathbb{V}^{E}_{L} = \mathbb{V}^{E}_{L,\mathrm{SS}}\mathbb{V}^{E}_{L,\mathrm{B}},
\end{equation}
where the seesaw angle is
\begin{equation}
\Theta^{E\dagger}_{L} =
\left(
\begin{array}{ccc}
 0 & \dfrac{h_{3e}^{ee}g_{\chi e}^{2e}+h_{3e}^{e\tau}g_{\chi e}^{2\tau}+h_{3\mathcal{E}}^{e2}g_{\chi \mathcal{E}}^{22}}{(g_{\chi e}^{2e})^2+(g_{\chi e}^{2\tau})^2+(g_{\chi \mathcal{E}}^{22})^2
 }\dfrac{v_{3}}{v_{\chi}} 
  e^{i \left(a_{E}-e_{E}\right)} & 0 \\
 0 & \dfrac{h_{3e}^{\mu\mu}g_{\psi e}^{2\mu}}{(g_{\chi e}^{2e})^2+(g_{\chi e}^{2\tau})^2+(g_{\chi \mathcal{E}}^{22})^2
 }
 e^{i \left(b_{E}-f_{E}\right)}  & 0 \\
 0 & \dfrac{h_{1e}^{\tau e}g_{\chi e}^{2e}+h_{1e}^{\tau\tau}g_{\chi e}^{2\tau}+h_{3\mathcal{E}}^{\tau 2}g_{\chi \mathcal{E}}^{22}}{(g_{\chi e}^{2e})^2+(g_{\chi e}^{2\tau})^2+(g_{\chi \mathcal{E}}^{22})^2
 }
  e^{i \left(c_{E}-e_{E}\right)} & 0 \\
\end{array}
\right),
\end{equation}
and $\mathbb{V}^{E}_{L,\mathrm{SM}}$, contained in $\mathbb{V}^{E}_{L,\mathrm{B}}$ has the mixing angles 

---------------------------------------------------------------

\begin{equation}
\label{eq:Electron-SM-Rotation-angles-ModelB}
\begin{split}
\tan \theta_{13}^{E,L} &\approx 
	-\frac{(h_{3e}^{ee}h_{1e}^{\tau e}+h_{3e}^{e\tau}h_{1e}^{\tau \tau}+h_{3\mathcal{E}}^{e2}h_{1\mathcal{E}}^{\tau 2})}
	{(h_{1e}^{\tau e}g_{\chi e}^{2e}+h_{1e}^{\tau \tau}g_{\chi e}^{2\tau}+h_{1\mathcal{E}}^{\tau 2}g_{\chi\mathcal{E}}^{22})}\frac{v_{3}}{v_{1}}e^{i (a_{E}-c_{E})}
	\\
\tan \theta_{23}^{E,L} &=0
	\\
\tan \theta_{12}^{E,L} &= 0 
\end{split}
\end{equation}
\end{widetext}

The exotic charged leptons $E^{1,2,3}$ have acquired mass at TeV scale. 
The heaviest SM lepton $\tau$ acquired a supressed mass at GeV scale through $v_{1}$ so as it does not acquire mass at hundreds of GeV, but at units of GeV. The lepton $\mu$ has acquired mass through $v_{3}$ without any suppression, so its mass remains at hundreds of MeV. Finally, the lightest lepton, $e$, got its mass through a Yukawa suppression. 

Summarizing, the FMH is achieved by the implementation of the VH together with the mass matrices obtained from the Yukawa Lagrangian, whose terms are constrained by the non-universal $\mathrm{U(1)}_{X}$ gauge and $\mathbb{Z}_{2}$ discrete symmetries. The fermion masses are outlined in the table \ref{tab:Summary-Fermion-Masses}.

\section{Discussion and Conclusions}
\label{sect:Conclusions}

\begin{table*}
\caption{Summary of fermion masses. }
\label{tab:Summary-Fermion-Masses}
\centering
\begin{tabular}{l|lc|lc}
Family	&	
Fermion	&			& 
Fermion	
	&		\\ \hline\hline
	&	\multicolumn{2}{c}{SM Up-like Quarks}
	&	\multicolumn{2}{c}{SM Down-like Quarks}	\\ \hline\hline	
1	&	$u$	
	&	$\dfrac{h_{u}^{2}-{h_{u}'}^{2}}{h_{t}}\dfrac{v_{3}}{\sqrt{2}}$
	&	$d$	
	&	$\dfrac{h_{d}^{2}-{h_{d}'}^{2}}{h_{J}^{1}}\dfrac{v_{3}}{\sqrt{2}}$ \\
2	&	$c$	
	&	$\dfrac{h_{c}^{2}-{h_{c}'}^{2}}{h_{T1}}\dfrac{v_{1}}{\sqrt{2}}$
	&	$s$	
	&	$\dfrac{h_{s}^{2}-{h_{s}'}^{2}}{h_{b}}\dfrac{v_{3}}{\sqrt{2}}$ \\
3	&	$t$	
	&	$\dfrac{h_{t}v_{1}}{\sqrt{2}}$
	&	$b$	
	&	$\dfrac{h_{b}v_{2}}{\sqrt{2}}$ \\ \hline\hline
	&	\multicolumn{2}{c}{SM Neutral Leptons}	
	&	\multicolumn{2}{c}{SM Charged Leptons}	\\ \hline\hline	
1	&	$\nu_{L}^{1}$	
	&	$\dfrac{\mu_{\mathcal{N}} v_{3}^{2}}{{\left(h_{{N}3}\right)}^{2}v_{\chi}^{2}}h_{\nu1}^{2}$
	&	$e$	
	&	$\dfrac{(h_{e}^{2}-{h_{e}'}^{2})h_{E2}-(h_{e}''^{2}-{h_{e}'''}^{2})h_{E2}'}{h_{\tau}^{2}-{h_{\tau}'}^{2}}\dfrac{v_{3}}{\sqrt{2}}$ \\	
2	&	$\nu_{L}^{2}$	
	&	$\dfrac{\mu_{\mathcal{N}} v_{1}^{2}}{{\left(h_{{N}2}\right)}^{2}v_{\chi}^{2}}h_{\nu2}^{2}$
	&	$\mu$	
	&	$\dfrac{h_{\mu}v_{2}}{\sqrt{2}}$ \\
3	&	$\nu_{L}^{3}$	
	&	$\dfrac{\mu_{\mathcal{N}} v_{1}^{2}}{{\left(h_{{N}1}\right)}^{2}v_{\chi}^{2}}h_{\nu3}^{2}$
	&	$\tau$	
	&	$\dfrac{h_{\tau}^{2}-{h_{\tau}'}^{2}}{h_{E2}}\dfrac{v_{1}}{\sqrt{2}}$ \\ \hline\hline


	&	\multicolumn{2}{c}{Exotic Up-like Quarks}
	&	\multicolumn{2}{c}{Exotic Down-like Quarks}	\\ \hline\hline	
1	&	$T^{1}$	
	&	$\dfrac{h_{T1}v_{\chi}}{\sqrt{2}}$
	&	$J^{1}$	
	&	$\dfrac{h_{J1}v_{\chi}}{\sqrt{2}}$ \\
2	&	$T^{2}$	
	&	$\dfrac{h_{T2}v_{\chi}}{\sqrt{2}}$
	&	$J^{2}$	
	&	$\dfrac{h_{J2}v_{\chi}}{\sqrt{2}}$ \\ \hline\hline
	&	\multicolumn{2}{c}{Exotic Neutral Leptons}	
	&	\multicolumn{2}{c}{Exotic Charged Leptons}	\\ \hline\hline	
1	&	$N_{R}^{1}$	
	&	$\dfrac{h_{N1}v_{\chi}}{\sqrt{2}}$
	&	$E^{1}$	
	&	$\dfrac{h_{E1}v_{\chi}}{\sqrt{2}}$ \\
2	&	$N_{R}^{2}$	
	&	$\dfrac{h_{N2}v_{\chi}}{\sqrt{2}}$
	&	$E^{2}$	
	&	$\dfrac{h_{E2}v_{\chi}}{\sqrt{2}}$ \\
3	&	$N_{R}^{3}$	
	&	$\dfrac{h_{N3}v_{\chi}}{\sqrt{2}}$ 
	&	$E^{3}$	
	&	$\dfrac{h_{E3}v_{\chi}}{\sqrt{2}}$ \\ \hline\hline	
\end{tabular}
\end{table*}

The article describes a nonuniversal abelian extension to the SM $\mathrm{G_{SM}}\otimes \mathrm{U(1)}_{X}$ to address FMH. The set of $X$ charges, shown in Tab. \ref{tab:Bosonic-content-A-B} is a solution of the chiral anomaly equations \eqref{eq:Chiral-anomalies} with different exotic sectors, composed by up-like $\mathcal{T}$ and down-like quarks $\mathcal{J}$, charged leptons $\mathcal{E}$ and Majorana fermions $\mathcal{N}$. The model  contains two $\mathcal{T}^{1,2}$, two $\mathcal{J}^{1,2}$, two $\mathcal{E}^{1,2,3}$ and three $\mathcal{N}_{R}^{1,2,3}$. All of them acquire masses at TeV scale through the VEV $v_{\chi}$. 


The $t$ quark acquires mass through $v_{1}$ without any kind of suppression, so its mass remains at hundreds of GeV. On the contrary, the $c$ quark mass turns out suppressed because of the SSM involving the $c$ quark and the exotic species $\mathcal{T}^{1}$ present in the mass matrix $\mathbb{M}_{U}$, given by the sub-block
\begin{align*}
\frac{1}{\sqrt{2}}
\begin{pmatrix}
h_{1u}^{21}v_{1}	&	
h_{1T}^{21}v_{1}	\\
g_{\chi u}^{12}v_{\chi}	&
g_{\chi T}^{11}v_{\chi}
\end{pmatrix},
\end{align*}
which yields the suppression of the $c$ quark mass
\begin{equation*}
\begin{split}
m_{c}^{2} &\approx  \frac{\left(h_{1u}^{21}g_{\chi u}^{12}-h_{1T}^{21}g_{\chi T}^{11}\right)^{2}}{(g_{\chi u}^{12})^{2}+(g_{\chi T}^{11})^{2}}
\frac{v_{1}^{2}}{2},
\end{split}
\end{equation*}
from hundreds to units of GeV since it acquires mass via $v_{1}$, in accordance with experimental observations. 
Similarly, the $u$ quark mass is suppressed by the SSM involving $u$ and $t$ in the following sub-block of the up-like quarks mass matrix
\begin{align*}
\frac{1}{\sqrt{2}}
\begin{pmatrix}
h_{3u}^{11}v_{3}	&	
h_{3u}^{13}v_{3}	\\	
h_{1u}^{31}v_{1}	&	
h_{1u}^{33}v_{1}
\end{pmatrix},
\end{align*}
obtaining the mass of the $u$ quark through $v_{3}$ with the subtraction of Yukawa coupling constants
\begin{equation*}
\begin{split}
m_{u}^{2} &\approx \frac{(h_{3u}^{11}h_{1u}^{33}-h_{3u}^{13}h_{1u}^{31})^{2}}{(h_{1u}^{31})^{2}+(h_{1u}^{33})^{2}}
\frac{v_{3}^{2}}{2},
\end{split}
\end{equation*}
lowering the mass from hundreds of to units of MeV. Moreover, the top quark obtains its mass at the scale of hundreds of GeV directly via $v_{1}$ with no suppression mechanism $$m_{t}^{2} \approx \left[(h_{1u}^{31})^{2}+(h_{1u}^{33})^{2}\right)]
\frac{v_{1}^{2}}{2}.$$

The $d$ quark obtains its mass through $v_{3}$ and an SSM with the exotic species $J^{1}$, 
\begin{align*}
\frac{1}{\sqrt{2}}
\begin{pmatrix}
h_{3d}^{11}v_{3}    &  
h_{3\mathcal{J}}^{11}v_{3}	\\
g_{\chi d}^{11}v_{\chi}	&	
g_{\chi \mathcal{J}}^{11}v_{\chi}	\\
\end{pmatrix},
\end{align*}
which gives the suppressed mass of the $d$ quark 
\begin{equation*}
\begin{split}
m_{d}^{2} &\approx \frac{(h_{3\mathcal{J}}^{11}g_{\chi d}^{11}-h_{3d}^{11}g_{\chi \mathcal{J}}^{11})^2}{(g_{\chi d}^{11})^2+(g_{\chi \mathcal{J}}^{11})^2}\frac{v_{3}^2}{2}.
\end{split}
\end{equation*}
so as the $d$ quark does not acquire mass at hundreds, but at units of MeV in accordance with phenomenological data. 
The model suppresses the mass of the $s$ quark with the $b$ quark because of the SSM
\begin{align*}
\frac{1}{\sqrt{2}}
\begin{pmatrix}
h_{3d}^{22}v_{3} 	&	h_{3d}^{23}v_{3}\\
h_{2d}^{32}v_{2}	&	h_{2d}^{33}v_{2}\\%
\end{pmatrix}
\end{align*}
yielding the mass eigenvalue of a light $s$ quark
\begin{equation*}
\begin{split}
m_{s}^{2}= &\frac{(h_{3d}^{23}h_{2 d}^{32}-h_{3d}^{22}h_{2 d}^{33})^2}{(h_{2 d}^{32})^2+(h_{2 d}^{33})^2}\frac{v_{3}^2}{2}.
\end{split}
\end{equation*}
Within this SSM, the bottom quark acquires its mass directly throught $v_{2}$
$$m_{b}^{2} \approx \left[(h_{2 d}^{32})^2+(h_{2 d}^{33})^2\right]\frac{v_{2}^{2}}{2}  $$

For the neutral sector, light active neutrinos and two-folded sterile heavy neutrinos at TeV scale are obtained with the employment of ISS. Moreover, the model selects the normal ordering, with the lightest active neutrino $\nu_{L}^{1}$ turning out massless when the smallest VEV $v_{3}$ is neglected. 

The model predicts the non-suppressed mass of the $\mu$ given by $m_{\mu} = h_{3 e}^{\mu\mu} v_{3}/\sqrt{2}$, which turns out at hundreds of MeV because of $v_{3}$. Furthermore, the charged lepton sector shows the largest SSM in the article involving $e$, $\tau$ and the exotic species $E^{2}$, 
\begin{align*} \frac{1}{\sqrt{2}}\begin{pmatrix}
h_{3e}^{ee}v_{3}	&	h_{3e}^{e\tau}v_{3}	&	h_{3\mathcal{E}}^{e2}3v_{3}	\\	
h_{1e}^{\tau e}v_{1}&	h_{1e}^{\tau \tau}v_{1}	&	h_{1\mathcal{E}}^{\tau 2}v_{1}	\\
g_{\chi e}^{2e}v_{\chi}	&	g_{\chi e}^{2\tau}v_{\chi}	&	g_{\chi \mathcal{E}}^{22}v_{\chi}	\\
\end{pmatrix}
\end{align*}
which gives the simultaneous suppression of the $e$ and $\tau$ masses, from hundreds to units of MeV and GeV, respectively. By setting $g_{\chi e}^{2e}$ null to simplify algebraic expressions, the masses of the $e$ and $\tau$ turn out to be
\begin{equation*}
\begin{split}
m_{ e  }^{2} &\approx  \frac{\left[(h_{3e}^{e\tau}h_{1e}^{\tau e}-h_{3e}^{ee}h_{1e}^{\tau \tau})g_{\chi \mathcal{E}}^{22}-(h_{3\mathcal{E}}^{e2}h_{1e}^{\tau e}-h_{3e}^{ee}h_{1\mathcal{E}}^{\tau 2})g_{\chi e}^{2\tau}\right]^2}{(h_{1\mathcal{E}}^{\tau 2}g_{\chi e}^{2\tau}-h_{1e}^{\tau \tau}g_{\chi\mathcal{E}}^{22})^2}\frac{v_{3}^2}{2},\\
m_{\tau}^{2} &\approx \frac{(h_{1\mathcal{E}}^{\tau 2}g_{\chi e}^{2\tau}-h_{1e}^{\tau \tau}g_{\chi\mathcal{E}}^{22})^2}{(g_{\chi e}^{2\tau})^2+(g_{\chi \mathcal{E}}^{22})^2}\frac{v_{1}^2}{2},
\end{split}
\end{equation*}
It is remarkable how the exotic $E^{2}$ suppresses the $\tau$ mass from hundreds to units of GeV, and in turn it suppresses the mass of the $e$ from hundreds to units of MeV, as it is shown in the above expressions. 

A summary of the fermion masses obtained from the model is shown in Tab. \ref{tab:Summary-Fermion-Masses}. 
The nonuniversal abelian extensions can be considered one of the simplest schemes beyond SM because it only comprises one abelian gauge group $\mathrm{U(1)}_{X}$. However, they give rich frameworks where fundamental issues such as fermion mass hierarchy can be addressed with the suited particle content and couplings. Moreover, this article shows how previous schemes \cite{mantilla2017nonuniversal} can be improved to avoid radiative corrections or fine-tunings and obtain in a natural way light and heavy fermions in accordance with experimental data. 

Regarding the global symmetries which appear from the assignation of $\mathrm{U(1)}_{X}$ quantum numbers in Table \ref{tab:Particle-content-A-B}, the discrete $\mathbb{Z}_{2}$ breaks them in the following scheme,
\begin{equation}
\begin{split}
    &\mathrm{SU(2)}_{q_{L}^{2,3}}\otimes
    \mathrm{SU(3)}_{u_{R}^{1,2,3}}\otimes
    \mathrm{SU(2)}_{d_{R}^{2,3}}
    \otimes
    \mathrm{SU(2)}_{ e _{R}^{1,3}}\\
    \overset{\mathbb{Z}_{2}}{\longrightarrow}&\;
    \mathrm{SU(2)}_{u_{R}^{1,3}}\otimes
    \mathrm{SU(2)}_{d_{R}^{2,3}}
    \otimes
    \mathrm{SU(2)}_{ e _{R}^{1,3}}, 
\end{split}
\end{equation}
which prevents the existence of massless fermions after the SSB. 

On the other hand, it is important to reinforce that the set of chiral anomaly-free $\mathrm{U(1)}_{X}$ quantum numbers constrains the set of $\mathrm{U(1)}_{X}$ quantum numbers in the scalar sector in order to understand the masses and mixing angles observed in the fermionic spectrum of the SM.

\section*{Acknowledgment}
This work was supported by \textit{El Patrimonio Autonomo Fondo Nacional de Financiamiento para la Ciencia, la Tecnolog\'{i}a y la 	Innovaci\'on Francisco Jos\'e de Caldas} programme of COLCIENCIAS in Colombia. 

\appendix 

\section{General scheme for diagonalization of mass matrices}
\label{sect:Appendix-Diag}

 The fermions of each sector are described in two bases: The flavor basis $\textbf{F}$ and the mass basis $\textbf{f}$. The mass matrix $\mathbb{M}_{F}$ writen in the basis $\textbf{f}$ is found in the diagonal form, having as diagonal elements the masses of the fermions.  By the implementation of spontaneous rupture of symmetry, the Yukawa Lagrangian $\mathcal{L}_{F}$ can be expressed as:
\begin{equation}\label{lagranfermi}
-\mathcal{L}_{F}=\overline{\textbf{F}_{L}}\mathbb{M}_{F}\textbf{F}_{R}+h.c.
\end{equation}
The diagonalization of the mass matrix is made by a biunitary transformation, which reads
\begin{equation}
(\mathbb{V}^{F}_{L})^{\dagger}\mathbb{M}_{F}\mathbb{V}^{F}_{R}.
\end{equation}
The unitary matrices $\mathbb{V}^{F}_{R}$ and $\mathbb{V}^{F}_{L}$ relate the mass and the flavor bases for both chiralities
\begin{align}
&\textbf{F}_{L}=\mathbb{V}_{L}^{F}\textbf{f}_{L}&\textbf{F}_{R}=\mathbb{V}_{R}^{F}\textbf{f}_{R}.
\end{align}
However, $\mathbb{M}_{F}$  is not a symmetric matrix and two unitary matrices have to be found. An alternative way is to diagonalize the symmetric constructions from  $\mathbb{M}_{F}$, which are:
\begin{align}
&\mathbb{M}_{F}\mathbb{M}_{F}^\dagger\rightarrow (\mathbb{V}^{F}_{L})^{\dagger}\mathbb{M}_{F}\mathbb{M}_{F}^\dagger \mathbb{V}^{F}_{L}\\ &\mathbb{M}_{F}^\dagger\mathbb{M}_{F}\rightarrow (\mathbb{V}^{F}_{R})^{\dagger}\mathbb{M}_{F}^\dagger\mathbb{M}_{F} \mathbb{V}^{F}_{R},
\end{align}
where the diagonal form is found only by the mixing of left-handed fermions, in the case of $\mathbb{M}_{F}\mathbb{M}_{F}^\dagger$, and by the mixing of the right-handed fermions, for the matrix $\mathbb{M}_{F}^\dagger\mathbb{M}_{F}$ (for neutrinos the mass matrix $\mathbb{M}_{N}$ is already symmertic.). Both quadratic mass matrices lead to the same eigenvalues, that correspond with the square of the masses. Since the main goal is to obtain the values of the fermion masses, only the matrix $\mathbb{M}_{F}\mathbb{M}_{F}^\dagger$ was diagonalized, giving also the Yukawa mixings of the left-handed fermions. The first step is to notice that the matrix $\mathcal{M}_{F}^{\mathrm{sym}}\equiv\mathbb{M}_{F}^\dagger\mathbb{M}_{F}$ can be written in the following block form:
\begin{equation}
\mathcal{M}_{F}^{\mathrm{sym}}=
\begin{pmatrix}\mathcal{M}_{3\times 3}^{f}&\mathcal{M}_{3\times n}^{f F}\\
\mathcal{M}_{n\times 3}^{Ff}&\mathcal{M}_{n\times n}^{ F}\end{pmatrix},    
\end{equation}
where $n$ is the number of exotic fermions and $(\mathcal{M}_{3\times n}^{f F})^T=\mathcal{M}_{n\times 3}^{Ff}$. Thus, a seesaw rotation was applied to separate the SM from the exotic sector\cite{grimus2001seesaw}
\begin{equation}
\mathbb{V}^{F}_{L,\mathrm{SS}}=\begin{pmatrix}
1&\Theta_{L}^{F\dagger}\\
-\Theta_{L}^{F}&1,
\end{pmatrix}    
\end{equation}
with $\Theta_{L}^{F}=(\mathcal{M}^{f})^{-1}\mathcal{M}^{Ff}$. This leaves the matrix in a block-diagonal form, as it was required
\begin{equation}
(\mathbb{V}^{F}_{L,\mathrm{SS}})^T\mathcal{M}_{F}^{\mathrm{sym}}\mathbb{V}^{F}_{L,\mathrm{SS}}=\begin{pmatrix}
m^{\mathrm{sym}}_{F,\mathrm{SM}}&0_{3\times n}\\
0_{n\times 3}&M^{sym}_{F,\mathrm{exot}}
\end{pmatrix},
\end{equation}
where $m^{sym}_{F,\mathrm{SM}}\approx \mathcal{M}^{f}+\mathcal{M}^{fF}(\mathcal{M}^{F})^{-1} \mathcal{M}^{fF}$ is the mass matrix for the SM and $M^{sym}_{F,exot}\approx \mathcal{M}^{F}$ the one for the exotic sector. 

Then, a diagonalization for each sector was performed, summarized in the following matrix:
\begin{equation}
\mathbb{V}_{\mathrm{B}}^{F}=\begin{pmatrix}
V_{\mathrm{SM}}^{F}&0_{3\times n}\\
0_{n\times 3}&V_{\mathrm{Exot}}^{F}
\end{pmatrix},    
\end{equation}
where $V_{\mathrm{SM}}^{F}$ is expressed by
\begin{equation}
V_{\mathrm{SM}}^{F}=R_{13}(\theta_{13}^F,\delta_{13}^{F})R_{23}(\theta_{23}^F,\delta_{23}^{F})R_{12}(\theta_{12}^F,\delta_{12}^{F}).    
\end{equation}
The matrices $R_{ij}$ are complex rotations, that read
\begin{equation}
\begin{split}
&R_{12}(\theta_{12}^F,\delta_{12}^{F})=\begin{pmatrix}
c_{12}^{F}&s_{12}^{F}&0\\
-s_{12}^{F*}&c_{12}^{F}&0\\
0&0&1
\end{pmatrix},\\
&R_{13}(\theta_{13}^F,\delta_{13}^{F})=\begin{pmatrix}
c_{13}^{F}&0&s_{13}^{F}\\
0&1&0\\
-s_{13}^{F*}&0&c_{13}^{F}
\end{pmatrix},\\
&R_{23}(\theta_{23}^F,\delta_{23}^{F})=\begin{pmatrix}
1&0&0\\
0&c_{23}^{F}&s_{23}^{F}\\
&-s_{23}^{F*}&c_{23}^{F}
\end{pmatrix},
\end{split}    
\end{equation}
where $c_{ij}^F=\cos{\theta_{ij}^{F}}$ and $s_{ij}^{F}=\sin{\theta_{ij}^{F}}\exp\left(i\delta_{ij}^{F}\right)$. The angles $\theta_{ij}^{F}$ are determined, in the calculus, from their tangents in an approximate way. Thus, the unitary transformation that leaves the symmetric mass matrix for the left-handed fermions in a diagonal form is
\begin{equation}
\mathbb{V}_{L}^{F}=\mathbb{V}_{L,\mathrm{SS}}^{F}\mathbb{V}_{L,\mathrm{B}}^{F}.    
\end{equation}

\vspace{0.1mm}

\end{document}